# TOWARDS BIOLOGY-ORIENTED TREATMENT PLANNING IN HADRONTHERAPY

Pavel Kundrát*
Institute of Physics, Academy of Sciences of the Czech Republic, Na Slovance 2, 182 21 Praha 8, Czech Republic

**By representing damage induction by ionizing particles and its repair by the cell, the probabilistic two-stage model provides a detailed description of the main processes involved in cell killing by radiation. To link this model with issues of interest in hadron radiotherapy, a simple Bragg peak model is used. Energy loss, its straggling, and the attenuation of the primary particle fluence are represented in a simplified way, based on semi-phenomenological formulas and energy-loss tables. An effective version of the radiobiological model, considering residual (unrepaired) lesions only, is joined with the simple physical model to estimate cell survival along ions' penetration depth. The predicted survival ratios for CHO cells irradiated by carbon ions are presented, showing very good agreement with experimental data.**

Existing hadrontherapy treatment planning approaches have represented physical processes to a great detail[1], while biological processes have been accounted for in a schematic manner only[2,3]. The need for developing detailed biology-oriented approaches has been addressed in the probabilistic two-stage model of radiobiological effects[4], which explicitly takes into account the interplay of DNA damage induction by ionizing particles and its repair by the cell. A simplified version of the model, considering lesions not repaired by the cell only (residual damage), has been applied successfully in studying the effects of single ion tracks by analyzing survival data for irradiation by mono-energetic protons and light ions[5,6]. In the present work, this effective scheme[5] is used together with a simple physical module representing light ions' Bragg peaks, with the aims to estimate the biological effects along penetration depth and make a step towards proposing a biology-oriented treatment planning approach; a more detailed method, representing repair processes explicitly, will be subject of future work.

## METHODS

The physical model starts from the energy-loss tables implemented in the SRIM-2003 code[7]. Energy-loss straggling is represented by a corresponding straggling of the actual penetration depth relative to that of a particle obeying the mean energy-loss characteristics; phenomenological range straggling formulas[8] are used in this step. The effect of nuclear reactions is included at the level of attenuating the primary particle fluence only, using nuclear interaction lengths $\lambda$ reported in Ref.[8]. The products of fragmentation reactions are not followed in the present approach. The effects of scattering phenomena are not reflected, either.

*Corresponding author: Pavel.Kundrat@fzu.cz

As a radiobiological component, the effective scheme based on the probabilistic two-stage model[4] but considering only damage not repaired by the cell[5] is used. The model has been generalized to correspond to Bragg peak irradiation conditions: At given depth $x$ for given incident beam energy $E$, the single-track $a(L)$ and combined damage probabilities $b(L)$ dependent on LET $L$, derived in Ref.[5], are weighted over LET spectra $\pi_{x,E}(L)$ generated by the Bragg peak model to estimate average damage probabilities per track,

$$a_{av}=\int a(L)\pi_{x,E}(L)dL, \quad b_{av}=\int b(L)\pi_{x,E}(L)dL. \quad (1)$$

Survival probability is calculated by (cf. Ref.[5])

$$S_{x,E,\Phi}=\Sigma_k(1-a_{av})^k[(1-b_{av})^k+kb_{av}(1-b_{av})^{k-1}]\exp(-h)h^k/k!, \quad (2)$$

representing both single-track and combined effects. The average number of tracks per cell nucleus (with cross section $\sigma$), $h=\sigma\Phi_0\exp(-x/\lambda)$, takes into account the reduction of primary particle fluence $\Phi_0$ due to nuclear reactions; the contribution of fragments to cell killing is neglected.

## RESULTS

In Figure 1, calculated Bragg peaks of 195 and 270 MeV/u carbon ions in water are compared to experimental data from Ref.[1], demonstrating that the simple physical model used here represents the main characteristics of experimental Bragg peaks. Since fragmentation processes are not taken into account, the model does not reproduce tails behind the peaks, and dose in plateau regions is slightly underestimated.

In Figure 2, model predictions for the survival of CHO cells along Bragg peaks of 195 and 270 MeV/u carbon ions are compared to experimental data from Refs.[9,10]. Based on the Bragg peak positions reported in these experiments (which differ from those of Ref.[1] shown in Figure 1), the incident energies were adjusted to 187 and 264 MeV/u. Input parameters of the calculations were the geometrical cross-section of CHO nuclei[11], $\sigma_{CHO} = 108$ μm$^2$, and parameters describing damage probabilities derived in Ref.[5] from analyzing survival data in mono-energetic setup[11].

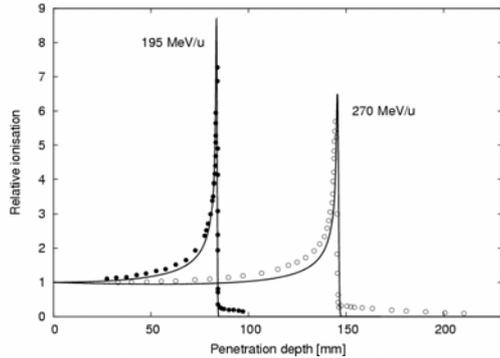

Figure 1: Bragg peak of carbon ions at 195 and 270 MeV/u. Model calculations compared to data from Ref.[1].

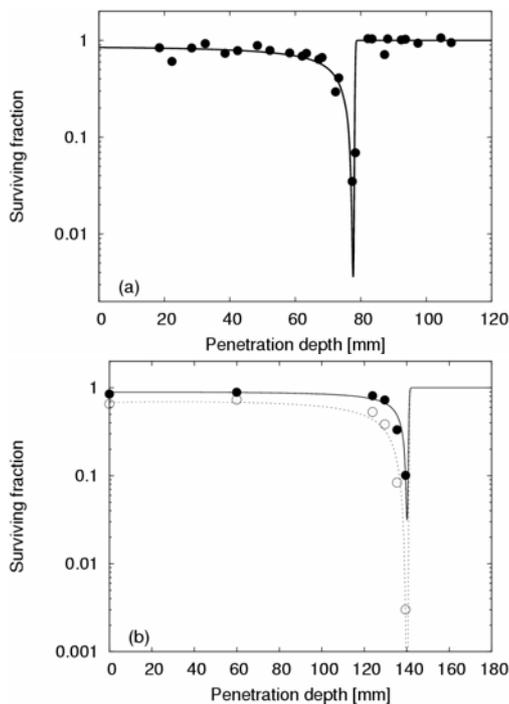

Figure 2: Survival of CHO cells irradiated by carbon ions with nominal energy 195 MeV/u, fluence $2 \times 10^7$ cm$^{-2}$ (panel a) and (panel b) 270 MeV/u, $2 \times 10^7$ cm$^{-2}$ (solid line, full symbols) and $5 \times 10^7$ cm$^{-2}$ (dotted line, empty symbols). Model predictions compared to experimental data[9,10].

DISCUSSION AND CONCLUSION

Although using a simple physical module, the model predicts survival along ions' penetration depth with a high precision necessary for treatment planning applications. At present, the predictive power of the model is limited by the use of mono-energetic survival data as input parameters to estimate the damage probabilities $a(L)$, $b(L)$. To overcome this issue, efforts will be made to relate these probabilities to the results of track structure studies. Attention will be devoted to the effects of fragments, too. Further generalization should also concern an explicit representation of repair processes, which at present are involved in residual damage characteristics; compare Refs.[4,12].

**Acknowledgment:** This work was supported by the grant "Modelling of radiobiological mechanism of protons and light ions in cells and tissues" (Czech Science Foundation, GACR 202/05/2728).